\documentclass[aps,prl,twocolumn,superscriptaddress,showpacs,amsmath,amssymb,longbibliography]{revtex4-1}
\usepackage[english]{babel}
\usepackage{amsmath}
\usepackage{bm}
\usepackage{graphicx,bbm} 
\usepackage{times}
\usepackage{epsfig} 
\usepackage[colorlinks,linkcolor=blue,citecolor=blue,urlcolor=blue]{hyperref}
\usepackage{color}
\usepackage{soul}
\definecolor{nred} {RGB}{224,0,0}
\definecolor{nblue} {RGB}{28,130,185}
\definecolor{dgreen} {RGB}{78,138,21}
\definecolor{norange}{RGB}{230,120,20}

\begin{document} 
\title{Complete many-body localization  in the  $t$--$J$ model caused by random magnetic field}
\author{Gal Lemut}
\affiliation{J. Stefan Institute, 1000 Ljubljana, Slovenia}
\author{Marcin Mierzejewski}
\affiliation{Department of Theoretical Physics, Faculty of Fundamental Problems of Technology, Wroc\l aw University of Science and Technology, 50-370 Wroc\l aw, Poland}
\author{Janez Bon\v ca}
\affiliation{Faculty of Mathematics and Physics, University of
Ljubljana, 1000 Ljubljana, Slovenia}
\affiliation{J. Stefan Institute, 1000 Ljubljana, Slovenia}
\begin{abstract} 
The many body localization (MBL) of  spin--$1/2$ fermions poses a challenging problem.  It is known that the disorder in the charge sector may be insufficient to cause full MBL. Here, we study dynamics of a single hole in one dimensional $t$--$J$ model subject to a random magnetic field. We show that strong disorder that couples only to the  spin sector localizes both spin and charge degrees of freedom. Charge localization is confirmed also for a finite concentration of holes. While we cannot precisely pinpoint the threshold disorder, we conjecture that there are two distinct transitions.  Weaker disorder first causes localization in the spin sector.  Carriers become localized for somewhat stronger disorder, when the spin localization length is of the order of a single lattice spacing. 
\end{abstract}
\maketitle

{\it Introduction.--} 
The many-body localization (MBL) has been 
demonstrated by various numerical \cite{ZZZ5_1,ZZZ5_2,ZZZ5_3,ZZZ5_4,ZZZ5_7,ZZZ5_8,ZZZ5_9,ZZZ5_10,ZZZ6_1,torres15,torres16} and analytical studies \cite{ZZZ5_5,ZZZ5_6} carried out 
mostly for one-dimensional (1D) systems of spinless particles or equivalent spin-models.  
Among 
unusual properties 
of MBL
we only emphasize the  logarithmic growth of the entanglement entropy \cite{znidaric08,bardarson12,kjall14,serbyn15,nc_luitz1,ZZZ4_1,ZZZ4_2}, and  
the subdiffusive transport in the regime of strong disorder but still below the MBL transition \cite{agarwal15,gopal15,znidaric16,lastsub}. 

While MBL is 
well understood for the simplest Hamiltonians, it is essential to recognize the class of more realistic quantum systems which may host this extraordinary phase. 
A challenging question concerns the dynamics of disordered two-dimensional interacting systems \cite{JJJ2,bordia2017_1, roeck2017} 
and  1D Hamiltonians which account for spin \cite{peter,mondaini15,gopal1,gopal2,bonca_2017,Hauschild_2016} or lattice degrees of freedom \cite{MOTT1968, emin1974} .  Numerical studies of the 1D Hubbard model \cite{mondaini15} suggest that the disorder strength needed for localization is  very large. 
Other results \cite{peter} obtained for the same model indicate that strong disorder in the charge sector localizes only charge carriers, while spin excitations remain delocalized. Similar studies carried out for the t-J model \cite{bonca_2017} 
 suggest that localization of theses carriers should be accompanied by localization of the spin degrees of freedom, otherwise the charge dynamics is subdiffusive up to the longest times accessible to the numerical calculations. 
 Such expectation may be supported also by the studies in Refs. \cite{gopal1,gopal2}.

	A general problem concerns the dynamics of a multicomponent system in the presence of disorder which couples exclusively to one of its subsystems. There is a quite convincing evidence that all subsystems \cite{MOTT1968,bonca_2017} or at least some of them \cite{peter} may be delocalized.  However, can such system show complete MBL where all degrees of freedom are localized?  In this work we show that it is indeed possible.  We consider a Hamiltonian, which is very similar to that in Ref. \cite{bonca2017}, namely, we study one-dimensional t-J model. However, the disorder is introduced not it the charge sector but in the spin sector through a random magnetic field \cite{herbrych2017} breaking the SU(2) symmetry  \cite{su21,su22,su23}. We show that such disorder may localize both charge and spin degrees of freedom.  We speculate also that there may be two localization transitions, one  for spin and the other for charge degrees of freedom.

{\it Model and method.} In the first part we investigate 
1D $t-J$ model with a single hole in a random external magnetic field $h_i \in [-W,W]$ 
\begin{equation}
H = -t_0\sum_{i,\sigma}\tilde c^{\dagger}_{i,\sigma}\tilde c_{i+1,\sigma} + c.c. +  J \sum_{i} \mathbf{S}_i\mathbf{S}_{i+1}  + \sum_i h_iS^z_i, \label{eq1}
\end{equation}
were $\tilde c_{i,\sigma} = (1-n_{i,-\sigma}) c_{i,\sigma}$ is a projected fermion operator. We perform calculations for 
various length sizes  $L$ and open boundary conditions. 
We perform time-evolution using Lanczos based technique. For most cases we use complete Hilbert spaces with a fixed total $S^z=0$. When computing the time evolution of the initially localized hole  we use the limited functional Hilbert space  (LFS) \cite{JJJ7,JJJ6,JJJ8,JJJ5}.
This method enabled calculations on larger chains up to a maximal size $L_\mathrm{max}=29$, described in more details in Ref.~\cite{supp}.

We start the time evolution from  a N\'eel background, with the hole  located  in the middle of the chain. In addition we compute static expectation values of various physical quantities for eigenstates in the middle of the energy band using ARPAC Lanczos techniques. We typically take $500$ realizations of the disorder. We measure time in units of $[1/t_0]$ and set $t_0=1$. If not specified otherwise, we  set also $J=1$.

 In order to investigate the dynamics of the charge carrier we calculate the hole  density 
\begin{equation}
\rho_i =\langle \psi |1-n_{i\uparrow}  -n_{i\downarrow} |\psi \rangle_\mathrm{ave},
\label{rho}
\end{equation}
where $\langle\rangle_\mathrm{ave}$ signifies that expectation values  have  been averaged over different random realizations of ${h_i}$. 
We also    define  the mean square deviation of the hole distribution \cite{robin2016}
\begin{equation}
\sigma^2=\sum_i i^2 \rho_i -\left[\sum_i i \rho_i \right]^2.
\label{vari}
\end{equation}

\begin{figure}[!htb]
\includegraphics[width=0.9\columnwidth]{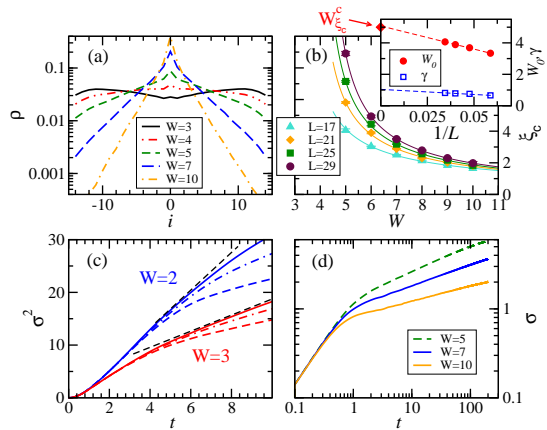}
\caption{ a)  The hole density $\rho_i$ at time $t=200$ for different values of $W$ as indicated in the insert.  The size of the system was $L=29$; b) extracted charge localization lengths $\xi_c$ for different system sizes vs. $W$. Thin lines represent fits of the form $\xi_c=A/(W-W_0)^\gamma$. Insert:  fit parameters extrapolated towards  $1/L=0$; c) $\sigma^2(t)$ for short times below the localization transition, $W=2$ and 3 showing diffusive behavior. Thin black dashed straight lines are guides to the eye. Dashed, dot-dashed and full lines represent systems sizes $L=21,25,$ and 29, respectively; d)   $\sigma(t)$ on $\log(t)$ scale for $W=5,7,$ and 10 for maximal system size $L=29$ using LFS.  }
\label{fig1}
\end{figure}

 Figure \ref{fig1}(a) shows  $\rho_i $ computed  at large time of evolution, {\it e.g.} $t=200$.  Note that for  $t=0$, the initial density is 
$\rho_i=\delta_{i0}$.
At small $W=2$ and 3 results are consistent with the delocalized state of the hole. 
In contrast, for $W\geq 5$,  $\rho_i$ is compatible with the localized state, 
$\rho_i \sim \exp(-|i|/\xi_\mathrm{c})$ for $i \ne 0$. Extracted charge localization lengths $\xi_c$ are presented in  Fig.~\ref{fig1}(b) for different system sizes $L$ as functions of $W$. Functional dependence of $\xi_c(W)$ can be well fitted using a divergent form as described  $\xi_c=A/(W-W_c)^\gamma$.  After $L\to \infty$ scaling we obtain a  threshold value $W^c_{\xi_c}\simeq 5$ separating delocalized regime (for $W<W^c_{\xi_c}$) from localized one.
Since the charge dynamics doesn't saturate for $t \le 200$, see Fig. \ref{fig1}d, while $\xi_c$ increases with time, we conclude that for $t\rightarrow \infty $ one gets $W^c_{\xi_c} \gtrsim 5$.

While the exponent $\gamma \simeq 1$ is consistent with  other results for spinless fermions (or equivalent spin model)  \cite{nc_luitz1, ZZZ5_3,lev15,nc_sirker}, it violates the so-called Harris-Chayes  bound  (HCB) $\gamma > 2 $ \cite{nc_chayes, nc_nand}. However, RG calculations predict a much larger $\gamma \approx 3.5$ \cite{vosk15,potter15}  consistent with the HCB. 
Violation of the HCB may originate from absence of a unique length--scale \cite{nc_sirker}.

We next follow the hole  dynamics via $\sigma^2(t)$. In Fig.~\ref{fig1}(c) we show short-time results for small values of $W=2$ and 3. We observe linear increase of $\sigma^2(t)$, consistent with the diffusive spread of the initially localized hole. At large values of $W=5,7,$ and 10  as shown in Fig.~\ref{fig1}(d),  we observe a subdiffusive propagation of the hole, $\sigma^2(t) \propto t^{\alpha}$ where the exponent $\alpha < 1$  decreases with increasing $W$. 
Eventually, for very large disorder, $\alpha$ becomes so small that the latter dependence is indistinguishable from the logarithmic increase of $\sigma(t)$    that is compatible with the proximity to the  MBL  state \cite{our2016}.

\begin{figure}[!htb]
\includegraphics[width=0.9\columnwidth]{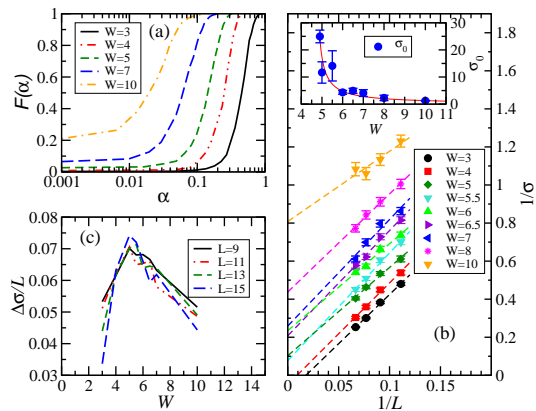}
\caption{ a) $F(\alpha)$  for different values of $W$. The largest available LFS Hilbert space with $L=29$ was used in this case; b) $1/\sigma$ scaling with the system size $L$. Insert: extrapolated values $\sigma_0$ (circles) with a fit (full line) on the functional form  $\sigma_0 \propto 1/(W-W^c_\sigma)^\gamma$ with $W^c_\sigma\simeq 5$ and $\gamma \simeq 0.95$ ; c) variance of $\sigma$ for different system sizes $L$.   Calculations in (b) and (c) were performed from eigenstates from the middle of the energy spectrum using  complete Hilbert spaces.  }
\label{fig2}
\end{figure}

Next, we check whether some particular realizations of disorder cause localization of the hole. We have thus fitted $\sigma^{2}(t) \propto t^{\alpha}$  independently for each realization of the disorder and obtained   the distribution of the exponents $f(\alpha)$.  We took special care to perform fits in the time domain free of finite-size effects. 
In Fig.~\ref{fig2}(a) we show the cumulative distribution function, 
\begin{equation}
F({\alpha})=\int_0^{\alpha} {\rm d} {\alpha'} f(\alpha'),
\label{cuma}
\end{equation}
We find  $F(\alpha \rightarrow 0)=0$ for $W < 5$ while  $F(\alpha \rightarrow 0)=F_0>0$ for $W=7$ and 10, which indicates localization.  

    We have also computed $\sigma$ for the case when $| \psi \rangle$ in Eq. (\ref{rho}) are eigenstates of the Hamiltonian taken from the middle of the energy spectrum.
 In Fig.~\ref{fig2}(b) we show $1/L$ scaling of $1/\sigma$.
 We can clearly see the transition from delocalized states where $1/\sigma(L\to\infty) \to 0$  for $W\lesssim 4.0$ towards localized ones with $1/\sigma(L\to\infty) \to 1/\sigma_0 >0 $ for $W\gtrsim 5.0$. In the inset we show scaling of extrapolated values $\sigma_0$ with $W$ together with a fit 
  $\sigma_0\propto 1/(W-W^c_\sigma)^\gamma$, which allows one to locate the divergence of $\sigma_0$ at $W^c_\sigma \simeq 5$.  Another signature of the MBL transition is observed in  variance (with respect
  to different realizations of disorder) of $\Delta\sigma/L$, presented in  Fig.~\ref{fig2}(c) that  shows a peak around $W\simeq 5$. Exactly at the transition we observe a linear scaling of  $\Delta\sigma$ with $L$ and, consequently,  $\Delta\sigma(W)/L$ becomes narrower as the system size increases. 

 The hole becomes localized at $W^c\simeq 5$ even though it is not directly subject to a random potential. 
We expect the localization of spin dynamics with increasing $W$  in the thermodynamic limit at the same value of $W^s\sim 3.7 \pm 0.5$ as in the undoped case \cite{pal10,ZZZ5_3,bera15},  since a single hole can not influence  the transition of an infinite chain. 
We test this idea by computing the entanglement entropy  $S=-\sum_\lambda w_\lambda \log w_\lambda$ where $w_\lambda$ are eigenvalues of the reduced density matrix of a subsystem. Since we work with odd system sizes, we have defined the reduced density matrix over a subsystem  of length $L_a=(L+1)/2$. While the subsystem contains spin as well as charge degrees of freedom, it is important to stress that there are only $L_a$  different states in the subsystem for the hole in contrast there is exponentially more spin degrees  of freedom.
In the thermodynamic  limit the entanglement entropy thus  measures predominantly   the entropy of the spin sector. 

\begin{figure}[!htb]
\includegraphics[width=0.9\columnwidth]{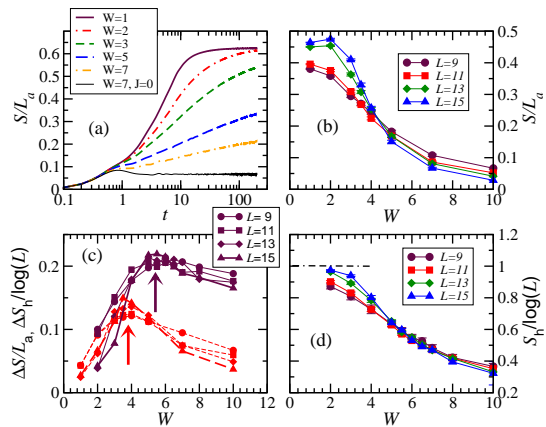}
\caption{ a)  $S/L_a$ for various values of $W$. Results were computed  using a complete basis on $L=13$ sites chain. Time evolution started from a N\' eel state with hole located in the middle of the chain. Thin black line represents Anderson's localized state for $W=7$ and  $J=0$;   b) $S/L_a$ computed from eigenstates from the middle of the energy band. The same holds for (c) and (d). Results are shown for various chain lengths, $L=9,11,13,$ and  $15$;  c) the variance of $S$ and $S_\mathrm{h}$ (symbols connected with dashed  and full lines respectively) vs. $W$; d) hole entropy $S_\mathrm{h}$  for different system sizes $L$.  }
\label{fig3}
\end{figure}

 The time evolution of the entanglement entropy shows a slow growth for $W\gtrsim 5$, {\it i.e.}   $S(t)/L \sim \log(t)$, as displayed in Fig.~\ref{fig3}(a), which is  consistent with the MBL state \cite{znidaric08,serbyn15}.  In contrast,  for small $W=1$ and 2, $S/L_a$ on a time scale $\tau\sim 10-50$ approach a constant slightly below  $\log(2)$,  which   represents infinite-$T$ limit of an undoped spin-one-half chain in a thermal state.   Transition between delocalized to localized regime can be well captured as well by following the size-dependence of the entanglement entropy $S/L_a$, \cite{ZZZ5_3}. In  Fig.~\ref{fig3}(b) we show $S/L_a$ vs. $W$ of the half-chain system obtained from eigenstates from the middle of the energy spectrum for different system sizes. We observe a crossover around $W^s\simeq 4$ as the system crosses over from the volume-law, characteristic for ergodic  and delocalized systems, towards the area-law, that signals localization as the sub-system size exceeds the localization length. In addition we show in Fig.~\ref{fig3}(c) the variance of the entanglement entropy $\Delta S/L_a$ that shows broad peak centered around $W^s\simeq 4$.

To gain additional insight into the localized phase 
 we trace out the spin degrees of freedom and obtain a reduced density matrix for 
the charge carrier. Consequently, the resulting von Neumann entropy, $S_\mathrm{h}$, quantifies entanglement between the spin  and the charge degrees of freedom. 
Deep in the MBL phase the charge and spin excitations are weakly entangled (see Fig. \ref{fig3}(d) ). Note also that the variance  $\Delta S_\mathrm{h}$ peaks at larger value of $W$ than $\Delta S$, see Fig.~\ref{fig3}(c).

Our results support  MBL at  large values of $W\gtrsim 5$ in the charge as well as in the spin sector. While MBL  in the spin sector is mostly expected based on many previous works \cite{znidaric08,pal10,ZZZ5_3,bera15}, the same is not true for the charge sector. 
An intuitive picture for the localization of the hole is obtained in the extreme anisotropic limit of the exchange interaction, {\it i.e.} in the limit when $J=J_z$ and even at  $J=0$.  Then, the system evolves within a space spanned by the states, $\vert\psi_i\rangle=|s_1, s_2,...,s_{i-1},0_i, s_{i+1},s_L \rangle $ with a frozen sequence (but not position) of $L-1$ spins $s_1,...,s_L$. As a result, 
the dynamics maps onto a problem of a single particle in a random on-site potential $\epsilon_i$ where 
\begin{eqnarray}
\epsilon_i &=& \sum_{j \ne i}h_j  s_j +J_z\sum_{j \ne i-1,i}  s_j s_{j+1}  
\end{eqnarray}
that is Anderson's localized  at  $W>0$. As an example we present data for $S/L_a(t)$ for $J=0$ in Fig.~\ref{fig3}(a) displaying rapid  saturation, characteristic for Anderson's localization. 
The picture of  frozen Ising--like spins is oversimplified in the presence of many--body interactions. It doesn't account for slow (logarithmic in time)
but  non-negligible  spin dynamics visible in Fig.~\ref{fig3}(a) for $J \ne 0$.  
Nevertheless, this results brings us to the hypothesis that the localization of the hole must be  caused by the localization 
of spin degrees of freedom.  We discuss this problem in more details at the  end of the manuscript as well as in the Supplemental Material \cite{supp}.

\begin{figure}[!htb]
\includegraphics[width=0.9\columnwidth]{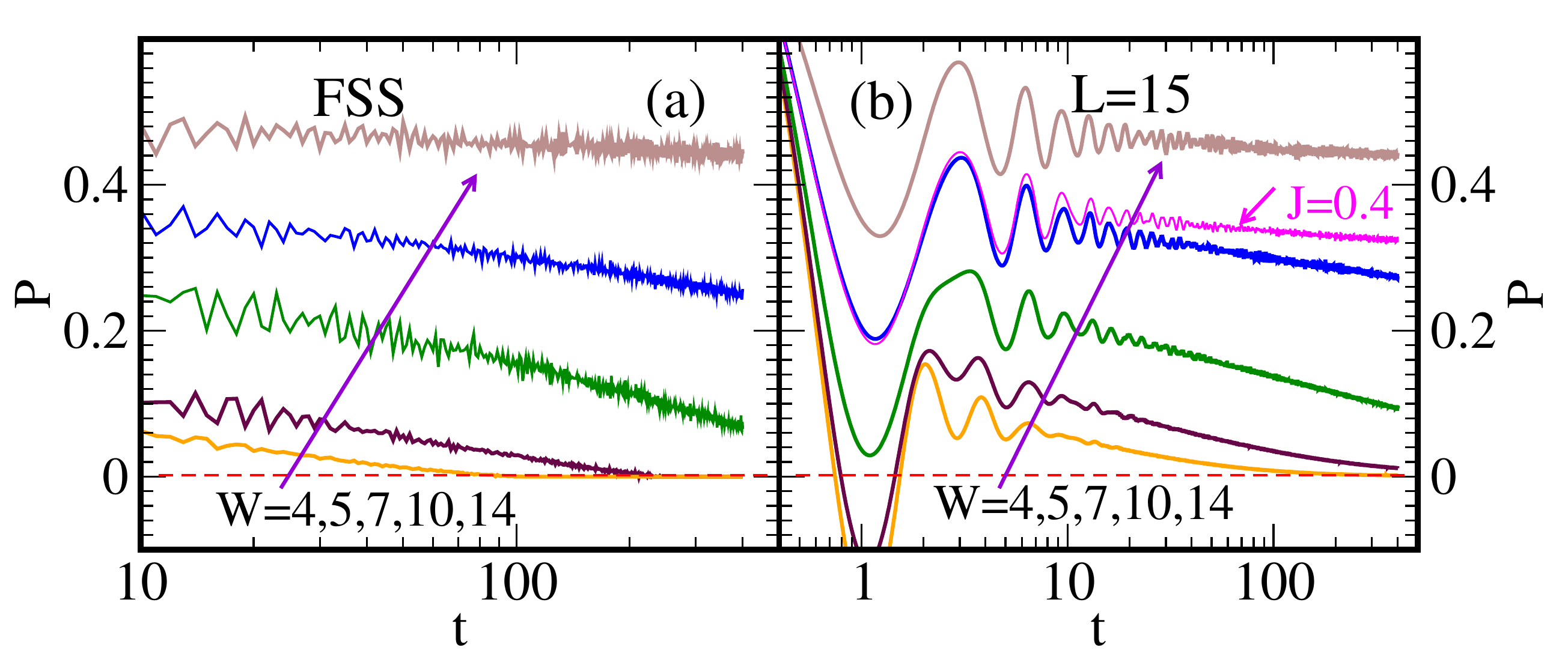}
\includegraphics[width=0.7\columnwidth]{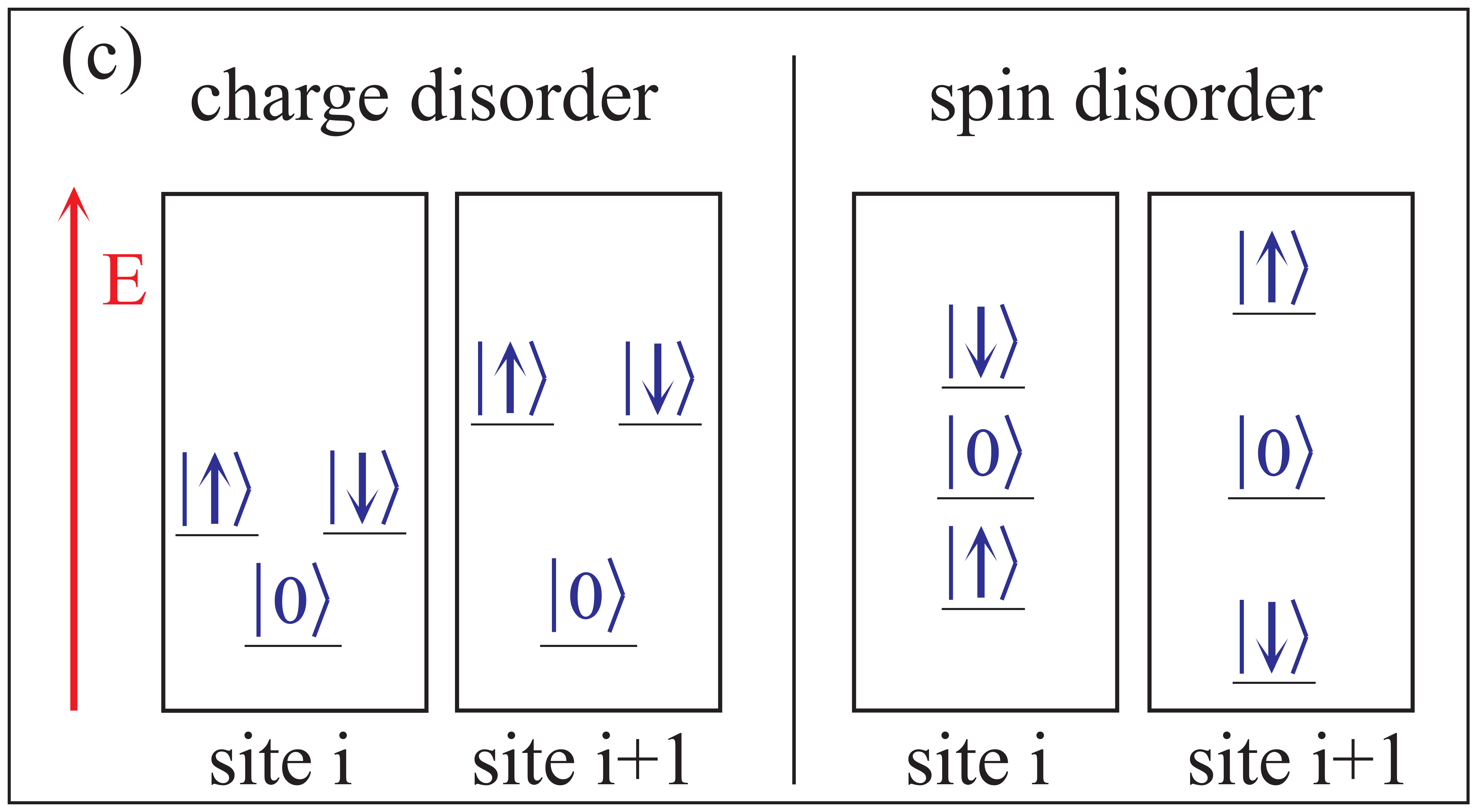}
\caption{ Time evolution of the charge imbalance  $P(t)$ of the  $t$-$J$ model with $L$ sites, $L/3$ holes and equal number of spin-up and spin-down Fermions. Results after finite size analysis are shown  in a); b) results at fixed $L=15$ and different values of $W$ compared with data for $J=0.4$ and $W=10$; c)  schematics portraying  diagonal energies of the basis states on neighboring sites for the case of charge and spin disorder.} 
\label{fig4a}
\end{figure}

{\it Finite doping.} The essential question is whether the randomness in the spin sector may induce the full MBL also for nonzero concentration of holes.
It is very demanding to carry out reliable finite--size scaling for arbitrary concentration of carriers.
A nontrivial but still numerically feasible case  concerns  the system with equal numbers  ($L/3)$
of holes,  spin--up and spin--down electrons.            
Following  Ref.  \cite{schreiber15}  we investigate the charge imbalance $P$. 
We study time evolution of initial states such that  every third lattice site (belonging to sublattice $A$) is occupied by holes whereas electrons are randomly distributed on the other sites which form the sublattice $B$.   Then, $P$ reads
\begin{equation}
P= \frac{3}{L}(\sum_{i \in A} \rho_i - \frac{1}{2} \sum_{i \in B} \rho_i).
\end{equation} 
The factor $1/2$ is the ratio of the number of sites in both sublattices. Initially all ($L/3$)  holes occupy the sublattice  $A$, hence $P(t=0)=1$.  Figure  \ref{fig4a}(b) shows 
 $P(t)$ where time propagation  has been carried out using full Hilbert space.   
Charge localization means that  the
system retains information on the initial distribution of holes for arbitrarily long times, i.e.,  $P(t \rightarrow \infty) > 0$.  This is clearly observed in Fig.~\ref{fig4a}(a) where at $W\gtrsim 10$ even after finite size analysis, Ref.~\cite{supp}, $P(t)$ displays slow logarithmic decay, characteristic for MBL, e.g.  see Ref. \cite{our2016}. 
In contrast, in the delocalized  phase  ($W\lesssim 5$) $P(t \rightarrow \infty) \rightarrow 0$ while it  starts to substantially deviate from 0 for  $W\gtrsim 7$, Fig.~\ref{fig4a}(b). In the latter figure we show also  results for smaller (more realistic) exchange interaction $J=0.4$, when the charge localization is even more evident. 

It is interesting that charge disorder is insufficient to induce full MBL\cite{bonca2017}, 
whereas random magnetic  field can localize all degrees of freedom. Most probably, this difference originates from a specific structure of  the Hilbert space which excludes double occupancy.  At each site,  the space is spanned by only three states $|\alpha_i\rangle$ with $\alpha=0,\uparrow,\downarrow$. The disorder in the charge-- and spin--sectors enter the Hamiltonian, respectively through terms
$
H'_{c,s}=\sum_i h_i  (|\uparrow_i \rangle \langle \uparrow_i| \pm |\downarrow_i \rangle \langle \downarrow_i |)
$,
with random $h_i$.
The basis states are eigenstates of $H'_{c,s}$, i.e.,  $H'_{c,s} |\alpha_i \rangle = E_{c,s} (\alpha)   |\alpha_i \rangle $. However, for  the charge disorder one finds degenerate eigenvalues $E_{c} (\uparrow)=E_c (\downarrow)$, whereas for spin disorder the spectrum $E_{s} (\alpha) $ is nondegenerate (see Fig.~\ref{fig4a}(c)).
In the case of spin disorder, the change of energy due to arbitrary rearrangement of spins or charges, $|\alpha_i \alpha'_j  \rangle  \langle \alpha'_i \alpha_j  |$ with $\alpha \ne \alpha'$,
is of the order of $W$.  Therefore, all degrees of freedom become localized for sufficiently strong disorder.  
However due to the degenerate spectrum obtained for charge disorder, the change of energy due to spin--flip  $|\uparrow_i  \downarrow_j \rangle \langle \downarrow_i \uparrow_j    | +H.c.$ 
is independent of $W$ and magnetic excitations may remain delocalized.

 {\it In summary} we have shown that a system with coupled charge and spin degrees of freedom may undergo a complete MBL transition due to  disorder which couples only to the spin sector.
Here, the complete MBL is understood as a phase where both charge and spin excitations are localized.   Support for this  conclusion comes from numerical studies of the  $t$-$J$ model in the low-doping regime and with random magnetic field. We have  carried out complementary studies of several quantities which consistently show for $J=1$ that the spin and charge degrees of freedom become localized when the magnitude
 of random field exceeds  $W^s \simeq 4$ and $W^c \simeq 5$, respectively. While the main purpose of this work is just to show existence of the complete MBL, we conclude that our results may be  consistent  with two separate transitions (or crossovers) at $W^s$ and $W^c>W^s$.   The charge degrees are not localized until the spin localization length is of the order of a single lattice spacing. However, due to the proximity of both transitions, this conjecture should be verified by additional numerical studies. 
While thorough numerical studies have been carried out for vanishing concentration of holes, we have shown that for sufficiently strong disorder full MBL  
 arises also for nonzero concentration of carriers.

 \acknowledgments
 J.B. acknowledges the financial support from the Slovenian Research Agency  (research core funding No. P1-0044) and M.M. acknowledges support by the project 2016/23/B/ST3/00647 of the National Science Centre, Poland.  This work was performed, in part, at the Center for Integrated Nanotechnologies, a U.S. Department of Energy, Office of Basic Energy Sciences user facility. 
\ \\
\ \\
\ \\
\centerline{{\bf \large Supplemental Material\\}}
\ \\
\setcounter{table}{0}
\makeatletter
\renewcommand{\theequation}{S\arabic{equation}}
\renewcommand{\thefigure}{S\arabic{figure}}

\section{Calculation of spin localization length $\xi_s$} 
\label{SLL}

\begin{figure}[!htb]
\includegraphics[width=0.9\columnwidth]{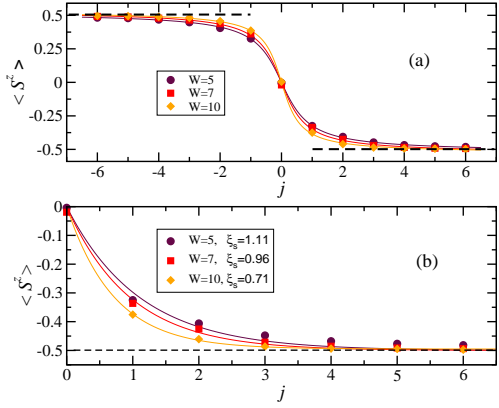}
\caption{ a)  $\langle S^z_j\rangle(t=200)$ for different values of $W>W^s$.  Full lines represent guides to the eye while dashed horizontal lines  represent values  of $\langle S^z_j\rangle(t=0)$; b) exponential fits with  a single parameter $\xi_s$: $<S^z_j>=-0.5(1-\exp(-j/\xi_s)$ for $j\geq 0$.  Results were obtained on a $L=13$ system with complete  Hilbert space. }
\label{figS1}
\end{figure}

In this section we show that for large enough $W$ spin localization length $\xi_s$ is smaller that the charge localization length $\xi_c$. To obtain an estimate of $\xi_s$ we follow ideas by J. Hauschild {\it et al} \cite{Hauschild_2016} where they computed  melting of a domain-wall in a spin-1/2 XXZ chain. In a similar fashion  we prepare our  system in a state where the left part of the chain ($j<0$) with $(L_a-1)-$sites has spins oriented up, the right-side ($j>0$) down,  while the hole is placed in the middle at $j=0$. We time evolve the system up to $t\sim 200$ and display results for $L=13$ using full Hilbert space in Fig.~\ref{figS1}. We observe  smearing  of the domain wall with an exponential decay of the spin redistribution penetrating  the respective domains. By fitting of  $\langle S^z_j\rangle$ with the exponential form,  as shown in Fig.~\ref{figS1}(b), we obtain $\xi_s=1.1,0.96$ and 0.71 for $W=5,7$ and 10  respectively, see Fig.~\ref{figS1}(b).  Comparing these results with $\xi_c$, presented in  Fig. \ref{fig1}(b) of the main text, we indeed obtain $\xi_c>\xi_s$.

\section{Finite-size scaling of 1/3 hole-doped $t$-$J$ model}
\label{FS}
Finite size scaling was performed on three different system sizes $L=9,12,$ and 15 with identical hole doping $n_\mathrm{hole}=1/3$. In all cases full Hilbert spaces and periodic boundary conditions were used. In the case of finite doping, we have studied a slightly modified version of $t$-$J$ model:
\begin{eqnarray}
H &=& -t_0\sum_{i,\sigma}\tilde c^{\dagger}_{i,\sigma}\tilde c_{i+1,\sigma} + c.c. \nonumber \\
   &+&  J \sum_{i} \left( \mathbf{S}_i\mathbf{S}_{i+1}  - n_in_{i+1}/4\right ) + \sum_i h_iS^z_i.
\end{eqnarray}
In the case of a single hole the term containing products of particle number operators  $n_i n_{i+1}$ represent only a constant shift of the energy. 

\begin{figure}[!htb]
\includegraphics[width=0.9\columnwidth]{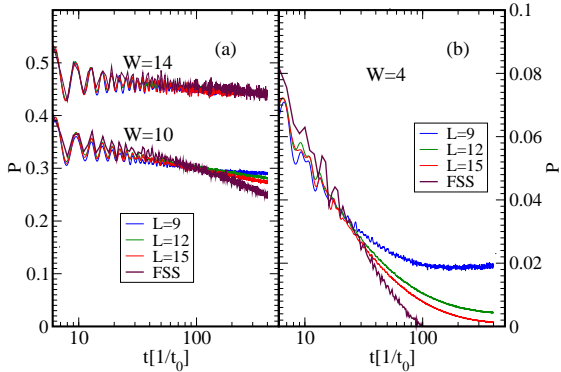}
\caption{ Charge imbalance $P$ (as defined in Eq.~(6) of the main text) vs. $t$ of the $t$-$J$ model with $L/3$ holes,  equal number of spin-up and spin-down fermions at different values of random field, a) $W=10$ and 14  and b) $W=4$. Different sizes $L$ are indicated in legends where in addition FSS indicates results where $1/L$ scaling was used separately for each point in time.  Results  in the case of $W=14$ are essentially independent on the system size. Parameters of the system are $t=J=1$. For time propagation  a time step $\Delta t = 0.02$ was used. Note also different vertical scales in a) and b). }
\label{figS2}
\end{figure}

\section{Limited Functional Hilbert Space for the $t$--$J$ model}  
\label{LFS}
Approach, described in this section is only used for the case of time propagation of a system with a single hole. Results, obtained using this approximate method are presented in Fig.~\ref{fig1} and Fig.~\ref{fig2}(a) of the main text. All other figures contain results obtained using complete Hilbert spaces. We first define  generators of the Limited Functional Hilbert Space (LFS) that are simply  off-diagonal parts of the Hamiltonian in Eq.~(\ref{eq1}) of the main text,
\begin{eqnarray}
O_1&=& \sum_{i,\sigma} c_{i+1\sigma}^{\dagger} c_{i\sigma} + H.c.  \nonumber \\ 
O_2&=& \sum_{i}  S^+_{i+1}  S^-_{i}+  S^-_{i+1}  S^+_{i} \label{hama}.
\end{eqnarray}
The generating algorithm starts from a hole at a given position, {\it e.g.} $i=0$ in a N\' eel state of spins, 
$\vert \psi^{(0)}\rangle = c_{0\sigma} \vert \mathrm{Neel}\rangle$.  We then apply the generator of basis $L$-times to generate the LFS:
\begin{equation} \label{lfs}
\left\{|\psi^{(l)} \rangle \right\}=\left( O_1 + \tilde O_2\right) ^{l}|\psi_{(0)} \rangle,
\end{equation}
for $l=0,...,L$. The operator $\tilde O_2$ acts only on pairs of spins that due to hole motion deviate from the original N\' eel state.  $L$ essentially represents the largest distance that the hole travels from its original position. In the case of LFS we impose open boundary conditions. After completing generation of LFS we time evolve the wave function using the Hamiltonian in  Eq.~(\ref{eq1})  of the main text  while taking the advantage of the  standard Lanczos-based diagonalization technique. Sizes of LFS span from $N_\mathrm{st}\sim 10^4$ for $L=21$ up to  $6 \times 10^4 $ for the largest $L=29$ used in our calculations. To achieve sufficient  accuracy of time propagation, we have used time-step-size  ranging from $\Delta t=0.1$ down to 0.02 and  performed up to $4\times 10^4 $ time steps. In addition we have sampled over $10^3$ different realizations of the random fields $h_i$. 

The main advantage of LFS over the exact diagonalization approach  is to significantly reduce the Hilbert space.  The  generation of  spin excitations is obtained  by the hole motion. The extent  of spin excitations away from the hole position is of the order of $L/2$ and it exceeds the spin localization length $\xi_s$ in the regime where  $W\gtrsim 5$  by one order of magnitude see also Fig.~\ref{figS1}.    Results for different quantities  computed using different  $L$ are presented in Figs.~\ref{fig1} (b) and (c) in the main text. The method has been successful in computing static and dynamic properties  \cite{JJJ8} as well as non-equilibrium dynamics \cite{JJJ7,JJJ6,JJJ5} of correlated electron systems.

\bibliography{references}
\end{document}